\documentclass[aps,prl,floats,twocolumn,showpacs,superscriptaddress]{revtex4}

\usepackage{epsfig}
\usepackage{latexsym,amsmath}

\begin{document}

\title{Conditional dynamics driving financial markets}

\author{Mari{\'a}n Bogu{\~n}{\'a}}

\affiliation{Departament de F{\'\i}sica Fonamental, Universitat de
  Barcelona, Avinguda Diagonal 647, 08028 Barcelona, Spain}
\email{mbogunya@ffn.ub.es}

\author{Jaume Masoliver}

\affiliation{Departament de F{\'\i}sica Fonamental, Universitat de
  Barcelona, Avinguda Diagonal 647, 08028 Barcelona, Spain}
\email{jaume.masoliver@ub.edu}

\date{\today}

\begin{abstract}
  We report empirical evidences on the existence of a conditional dynamics
  driving the evolution of financial assets which is found in several
  markets around the world and for different historical periods. In
  particular, we have   analyzed the DJIA database from 1900 to 2002
  as well as more than 50 companies trading in the LIFFE market of
  futures and 12 of the major European and American treasury bonds. In
  all of the above cases, we find a double dynamics driving the financial
  evolution depending on whether the previous price went up or down. We
  conjecture that this effect is universal and intrinsic to all markets
  and, thus, it could be included as a new stylized fact of the market.
\end{abstract}


\maketitle

One fundamental assumption lying behind many modern theories of
mathematical finance is the so called ``efficient market
hypothesis" which basically states that the market incorporates
instantaneously any information concerning future market evolution
\cite{fama1}. In consequence, if a market is efficient with
respect to some information set it is impossible to make economic
profits by trading on the basis of that information set
\cite{campbell-lo}. Observe that this in particular indicates that
market efficiency necessarily implies the absence of
(auto)correlations in financial prices at any time scale, for
correlation means some degree of predictability which in turn
would open the door to profitable strategies exclusively based on
the information contained in the price itself. Note incidentally
that the lack of correlations implied by the efficient market
hypothesis means that the price process must be driven by white
noise. However, this assumption is very restrictive since real
markets are not efficient, at least at short times, and the
existence of correlations seems to be well documented
\cite{campbell-lo,fama2,Mantegna_Stanley,Bouchaud_Potters}.

Therefore, the search for correlations in financial time series
has been the subject of intense research during the last years
\cite{campbell-lo,fama2,Mantegna_Stanley,Bouchaud_Potters,Ding93,Liu99,Gopi99,Potters98}.
Partly due to the hope that this knowledge would be useful for
predicting the behavior of the market and, in a more academic
sense, because such correlations could bring some light to the
understanding of real markets.

A good example of such correlations appears when using high
frequency data, that is, tick-by-tick data corresponding to the
evolution of a given asset. Thus, it has been observed that the
logarithmic variations of the price --the so called returns-- are
correlated with themselves in such a way that highly positive
returns are followed by also highly positive returns, and vice
versa. This means that the behavior of the price at some time
certainly depends on the past variations of the signal. However,
this correlation is found to be of short range, indeed, the
influence that a given variation of price has on the future
evolution decays exponentially, with a characteristic time of the
order of minutes \cite{Liu99,Gopi99}.

Another paradigmatic example is obtained by using daily variations of
the price. In this case the magnitude of interest is not
the return itself but its absolute value (or its square), which is
a measure of the fluctuations of the return or, in economic
terminology, the market volatility \cite{Liu99,Gopi99}. In this
case, an analysis of the correlation function shows a clear
positive correlation between the absolute value of the return
today and in the future, with a characteristic time of the order
of years. This persistent correlation is the responsible for
long periods of extremely volatile markets followed by calm periods,
that is, for the clusterization of the volatility.

A third example is provided by the leverage effect
\cite{Black76,Bouchaud01,Perello03}, which states that a large
drop of the price is followed by an increase of the volatility.
This correlation is found to be of intermediate range, with a
typical time scale of $10 \sim 50$ days. Summarizing, there
exist return-return correlations at very short times
(minutes), volatility-volatility correlations at very long
times (years) and return-volatility correlations at an
intermediate time scale (days). At the light of these results, one
may wonder whether it would be possible to find return-return
correlations at other temporal scales, for instance, at daily
scales.

In order to answer this question we will analyze a very simple
problem. Let us consider the temporal series of the sign of the
daily price returns \footnote{If $S_n$ and $S_{n-1}$ are the
closing market prices corresponding to the days $n$ and $n-1$
respectively, we define the daily price return of the day $n$,
$R_n$, as $R_n=S_n/S_{n-1}-1$. }, that is, $+1$ if a given day had
a positive increment of price and $-1$ otherwise. Which are the
statistical properties of this signal? The simplest hypothesis is
to consider that positive and negative days are uncorrelated
random events --as far as the sign is concerned-- with a
probability $p_+$ for positive days and $1-p_+$ for negative ones.
Using this  ansatz as a test or ``null hypothesis", the
probability of having a sequence of $n$ consecutive positive
returns, $\psi_+(n)$, is
\begin{equation}
\psi_+(n)=p_+^{n-1}(1-p_+), \label{psis}
\end{equation}
that is, an exponential law (Poisson law). The average number of
consecutive positive days is simply given by
$\langle\tau_+\rangle =(1-p_+)^{-1}$ and the same holds for sequences of days
with negative returns replacing $p_+$ by $1-p_+$, {\it i. e.}, $\langle\tau_-\rangle =p_+^{-1}$.

\begin{figure}
\epsfig{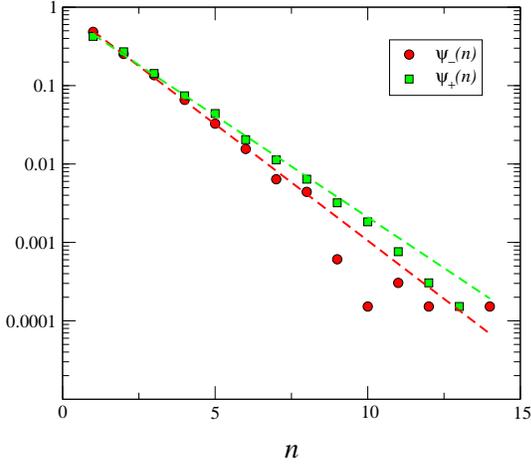} \caption{Probability of having a
sequence of $n$ consecutive positive or negative days. The solid
lines are the Poisson distributions discussed in the text with
average values given by $\langle \tau_+ \rangle_{empiric}=2.22$
and $\langle \tau_- \rangle_{empiric}=2.02$.}
 \label{psi}
\end{figure}

In order to accept or reject the null hypothesis, we use data from
the Dow Jones Industrial Average index (DJIA), which contains
daily records from 1900 to 2002 ($28126$ days), thus covering a
wide temporal range with many different economic and political
situations and providing a large database. Direct measurements on
this database yield for the frequency of positive days the value
$p_+=0.522 \pm0.002$ and, according to the model given by Eq.
(\ref{psis}), the expected number of consecutive positive and
negative days is $\langle \tau_+ \rangle_{theoretical}=2.09 \pm
0.01$ and $\langle \tau_- \rangle_{theoretical}=1.91 \pm 0.01$
respectively. Figure \ref{psi} shows the probability distributions
of the lengths of sequences of positive and negative days,
$\psi_+(n)$ and $\psi_-(n)$ . As it is clearly seen, these
distributions follow an exponential law, in agreement with
Eq.~(\ref{psis}). However, the empirical average lengths of
positive and negative days obtained from direct measurements are
$\langle \tau_+ \rangle_{empiric}=2.22 \pm 0.02$ and $\langle
\tau_- \rangle_{empiric}=2.02 \pm 0.02$ which are higher than the
theoretical values predicted above. The disagreement between
empiric and theoretical results is certainly small and might go
easily unnoticed, although a careful analysis of the statistical
errors leads to the rejection of the original null hypothesis on
the independence of positive and negative returns.

These results seem to point out that the market behaves
differently whenever there is a sequence of positive or negative
returns. On the other hand, the Poisson form for the distribution
of lengths of those sequences indicates that the market is
Markovian. Thus, no information can be extracted from the elapsed
time since the last change of sequence and, therefore, the memory
of the market must be, at most, of one single day. This implies
that the return of the price during a given day can only be correlated
with the previous day, in particular with the sign of the previous
day.

The simplest model able to reproduce all these empirical
observations is a two-state model, in which the probability of
having a positive or negative return depends on the sign of the
previous day. More precisely, let $p_{++}$ be the probability of
having a positive return given that the return of the previous day
was positive and $p_{--}$ the probability of having a negative
return given that the return of the previous day was negative.
Notice that the model has only two independent parameters, $p_{++}$ and
$p_{--}$, and the rest of probabilities can be obtained from them
as $p_{-+}=1-p_{++}$ and $p_{+-}=1-p_{--}$, measuring the
probability of having a negative (positive) return given that the
return of the previous day was positive (negative).

Using all these ingredients, the distributions $\psi_+(n)$ and
$\psi_-(n)$ are now given by
\begin{equation}
\psi_+(n)=p_{++}^{n-1} (1-p_{++})
\end{equation}
and
\begin{equation}
\psi_-(n)=p_{--}^{n-1} (1-p_{--}).
\end{equation}
Again a Poisson distribution with average length given by
$\langle \tau_+ \rangle=(1-p_{++})^{-1}$ and $\langle \tau_-
\rangle=(1-p_{--})^{-1}$ respectively. Finally, the frequency of
positive days, $p_+$, is \cite{Marian00}
\begin{equation}
p_+=\frac{1-p_{--}}{2-p_{--}-p_{++}},
\end{equation}
and a similar expression for $p_-$ exchanging  $p_{++}$ by $p_{--}$.

We will now compare the predictions of the two-state model with
empirical data. Direct measurements on the DJIA dataset yield the
following values for the conditioned probabilities: $p_{++}=0.547
\pm 0.004$ and $p_{--}=0.495 \pm 0.004$. Using these two measures
as inputs for the model, the predicted values for $p_+$, $\langle
\tau_+ \rangle$, and $\langle \tau_- \rangle$ are
$p_+=0.527\pm0.004$, $\langle \tau_+ \rangle=2.21\pm0.02$ and
$\langle \tau_- \rangle=1.98 \pm 0.02$, in perfect agreement with
the empirical results reported above.

\begin{widetext}

\begin{table}
\caption{Summary of the empiric statistics for the DJIA index
compared to the predictions of the uncorrelated model and the
two-state model. The empty values in rows 3 and 4 are taken from
the empiric measurements in row 2 as the inputs for the
theoretical predictions of each model. Numbers within parentheses are
affected by statistical errors.}
\begin{tabular}{l c c c c c c c c}
\hline \hline
   & $p_+$ \hspace{0.1cm}& $p_{++}$ \hspace{0.1cm}& $p_{--}$ \hspace{0.1cm}& $\langle \tau_+ \rangle$ \hspace{0.1cm}& $\langle \tau_- \rangle$ \hspace{0.1cm}& $\langle R_+ \rangle$ \hspace{0.1cm}& $\langle R_-
   \rangle$ \hspace{0.1cm}& $\langle R \rangle$
   \\[0.1cm] \hline
  Empiric DJIA index \hspace{0.2cm} & $0.52(2)$ \hspace{0.1cm}  & $0.54(7)$ \hspace{0.1cm}& $0.49(5)$ \hspace{0.1cm}& $2.2(2)$ \hspace{0.1cm}& $2.0(2)$ \hspace{0.1cm}& $8.(6)\cdot 10^{-4}$ \hspace{0.1cm}& $-4.(4)\cdot 10^{-4}$ \hspace{0.1cm}& $2.(4)\cdot 10^{-4}$\\
  Uncorrelated model \hspace{0.2cm}  & -- \hspace{0.1cm} &  $0.52(2)$ \hspace{0.1cm} & $0.47(8)$ \hspace{0.1cm} & $2.0(9)$ \hspace{0.1cm} & $1.9(1)$ \hspace{0.1cm}& $2.(4)\cdot 10^{-4}$ \hspace{0.1cm} & $2.(4)\cdot 10^{-4}$ \hspace{0.1cm} & --\\
  Two-state model  \hspace{0.2cm} & $0.52(7)$ \hspace{0.1cm} & -- & -- & $2.2(1)$ \hspace{0.1cm} & $1.9(8)$ \hspace{0.1cm} & --  \hspace{0.1cm}& -- \hspace{0.1cm} & $2.(5)\cdot 10^{-4}$ \\
  \hline \hline
\end{tabular} \label{table1}
\end{table}

\end{widetext}

It might be argued that the discrepancy between the empirical
measures of $\langle \tau_- \rangle$ and $\langle \tau_+ \rangle$
and the theoretical predictions of the uncorrelated model are
marginal and, consequently, the two-state model only introduces a
slight correction to the actual dynamics. However, what seems to
be significant is that the market apparently reacts differently
depending on the sign of the previous day, which naturally
introduces the idea of a double dynamics. Having this in mind, we
define $P(R|R_{prev}>0)dR$ to be the conditional probability that
the daily return lies within the interval $(R,R+dR)$ given that
the previous day had a positive return. Analogously $P(R
|R_{prev}<0)dR$ is that conditional probability if the previous
day had a negative return. Up to this point, we have only studied
the behavior of the sign of the signal specified by the quantities
$p_{++}$ and $p_{--}$, which are related to the previous functions
by
\begin{equation}
p_{++}=\int_0^{\infty} P(R |R_{prev}>0)dR
\end{equation}
\begin{equation}
p_{--}=\int_{-\infty}^{0} P(R |R_{prev}<0)dR.
\end{equation}
However, if the market is really driven by a double dynamics there
should be a substantial difference between the moments of $P(R
|R_{prev}>0)$ and $P(R |R_{prev}<0)$. Let us denote by $\langle
R_+ \rangle$ and $\langle R_- \rangle$ the first moment of these
distributions, that is, $\langle R_+\rangle$ [$\langle
R_-\rangle$] is the conditional average of the daily return given
that yesterday's return was positive [negative]. Let us denote by
$\sigma_+$, and $\sigma_-$ their standard deviation. For the DJIA
index, the empirical values of these quantities  are: $\langle R_+
\rangle=(8.6\pm0.8)\times 10^{-4}$, $\langle R_-\rangle=(-4.4\pm
1.0)\times 10^{-4}$, $\sigma_+=(9.9\pm 0.2)\times 10^{-3}$, and
$\sigma_-=(11.6\pm 0.3)\times 10^{-3}$. These values should be
compared with the unconditional average of the daily return,
$\langle R \rangle=(2.39 \pm 0.6)\times 10^{-4}$, and volatility
$\sigma=(10.7\pm 0.2)\times 10^{-3}$. Note that $\langle R
\rangle$ and its variance can be evaluated through the two-state
model by
$$
\langle R \rangle =p_+ \langle R_+ \rangle +p_- \langle R_-
\rangle,
$$
and
$$
\sigma=\sqrt{p_+\sigma_+^2 +p_-\sigma_-^2+p_+p_-
(\langle R_+\rangle-\langle R_-\rangle)^2},
$$
with the result $\langle R \rangle=(2.49 \pm 0.6)\times 10^{-4}$
and $\sigma=(10.7\pm 0.2)\times 10^{-3}$. Both in very good
agreement with their empirical values. Table \ref{table1}
summarizes the relevant statistics for the DJIA index and the
equivalent values predicted by the uncorrelated model and the
two-state model.

There is something quite remarkable in these results, since they
show that the average return of the market is the result of the
composition of two independent signals: one of them positive,
$\langle R_+ \rangle$, and another one negative, $\langle R_-
\rangle$. At the light of these results, and given the
multiplicative character of the market, it seems not to be
possible to neglect the effects of this double dynamics, at least
in the long run. Indeed, the quantitative difference between the
average daily return of both signals is rather significant in the
sense that a small change in the signal would substantially alter
the long term trend of the market.

\begin{figure}
\epsfig{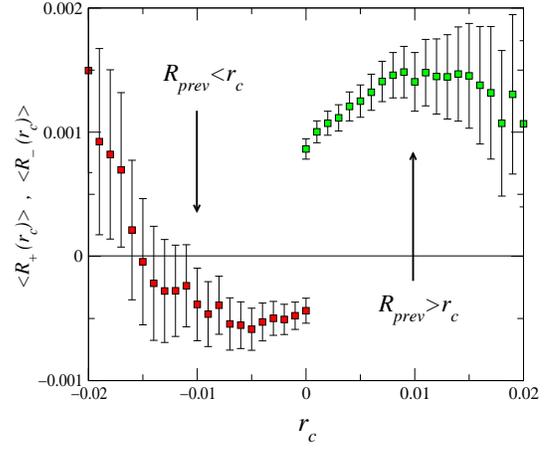} \caption{Average daily return
given that the previous day had a return greater than $r_c$
(right) and given that the previous day had a return smaller than
$r_c$ (left).}
 \label{retcond}
\end{figure}

As we have seen, the daily return of a given day is a random
quantity correlated with the return of the previous day. One
question that arises now is: how does this correlation depend on
the magnitude of the previous return. In order to check this
point, we have calculated the average return given that the
previous day had a return greater than a certain value $r_c$, $\langle
R_+(r_c)\rangle$, or smaller than $r_c$, $\langle R_-(r_c)\rangle$.
Notice that $r_c=0$ correspond to the previous analysis. These two
functions are plotted in Fig.\ref{retcond}. As is clearly seen,
there is a significant difference whenever the previous day had a
positive or negative increment in price. Thus for $r_c \in
[-1.5\%,1.5\%]$ the positive branch is positive --and slightly
increasing-- whereas the negative branch remains negative. Beyond
this interval, the negative branch increases and, eventually, both
branches become equivalent --given the statistical errors--
indicating that correlations are lost for this range of
returns, in other words, there is no net effect if the previous day had a
return greater than $1.5\%$ or smaller than $-1.5\%$. However,
these days represent less than $10\%$ of the total of trading days and they
do not lessen the relevance of the correlations present in the remainder
$90\%$ of trading days.

\begin{figure}
\epsfig{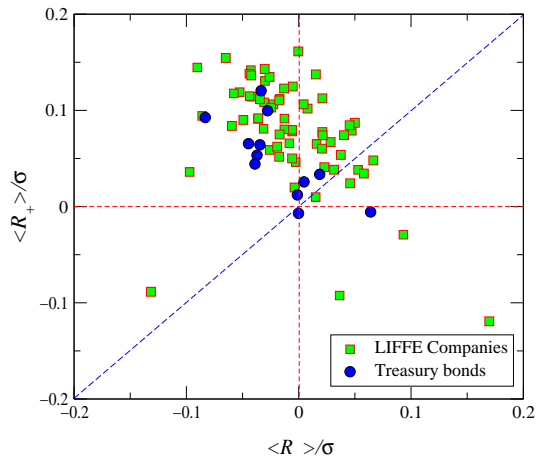} \caption{Scattered plot of
the average daily returns given that the previous day had a
positive or negative increment for the companies trading in the
LIFFE market and several European and American treasury bonds
during the period 1990-2002. For the sake of comparison, these
average returns are rescaled by the volatility of the
corresponding company.}
 \label{liffe}
\end{figure}

Finally, we address the question of universality. The preceding
analysis has been carried out for one specific index, the DJIA,
during a period of $100$ years, and the question is whether this correlation
is also present for individual stocks and any other class of financial
assets. In order to shed some light on this question, we have
analyzed the performance of more than $50$ companies trading in
the LIFFE market \footnote{London International Financial Futures and Options Exchange}
during the period 1990-2002. For each company, we have measured $\langle R_+
\rangle$ and $\langle R_- \rangle$. The results are shown in Fig.
\ref{liffe} as a scattered plot, with each axis representing each
of the conditioned average daily returns rescaled by the unconditional
volatility of the corresponding company. If no
correlation were present between the return and the sign of the
previous day then $\langle R_+ \rangle$ and $\langle R_- \rangle$
would take the same value (except for statistical fluctuations)
and, therefore, all companies would be scattered around the main
diagonal, in the first quadrant. In contrast we see if Fig. \ref{liffe} that
there is a clear tendency to stay in the second quadrant, with $\langle R_+
\rangle$ being a positive quantity and $\langle R_- \rangle$ being
a negative one (or close to zero). This means that, on average,
the returns after a positive day outperform those after a negative
one in agreement with the model presented. The same effect is found
in other classes of financial assets, such as treasury bonds
(blue symbols in Fig. \ref{liffe}) or commodities (not reported here).
All these results suggest, indeed, the universality of this double dynamics
driving the evolution of financial markets.

In summary, we have reported empirical evidences of the existence
of a conditional dynamics driving the behavior of financial
markets which manifests itself in the fact that daily prices tend
to go up or down depending on whether yesterday's price went up or
down. Moreover this dynamics seems to be ubiquitous to a wide
sample of different markets which may indicate the universal
character of this effect. We finally stress the fact that
financial time series are often non-stationary, at least at long
times, and, consequently, it is possible to find short periods in
which the double dynamics is not clearly visible. Therefore, the
empirical findings reported here must be considered from an
overall point of view at the same level as the observation that
the market is historically growing despite the existence of many
bear periods. This point will be addressed in future
communications.

\begin{acknowledgments}

We greatly acknowledge J. M. Porr\`{a}, J. -P. Bouchaud, and R.
Pastor-Satorras for many discussions and useful suggestions. This
work has been partially supported by DGI under Grant No.
BFM2000-0795 and by GC under contract No. 2000 SGR-00023. M. B.
acknowledges support from the European Commission FET Open Project
No. COSIN IST-2001-33555.

\end{acknowledgments}

\end{document}